\documentclass[journal=jacsat,manuscript=article]{achemso}

\usepackage[version=3]{mhchem} 
\usepackage[T1]{fontenc}       
\usepackage{indentfirst}
\usepackage[sort&compress,numbers,super]{natbib}  
\usepackage{achemso}
\usepackage{caption}
\usepackage{subcaption}
\usepackage{amssymb}

\author{Jaeyun Moon}
\affiliation{%
	Division of Engineering and Applied Science\\
	California Institute of Technology, Pasadena, California 91125,USA
}%
\author{Austin J. Minnich}%

 \email{aminnich@caltech.edu}
\affiliation{%
 Division of Engineering and Applied Science\\
 California Institute of Technology, Pasadena, California 91125,USA
}%

\title[An \textsf{achemso} demo]
  {Sub-amorphous thermal conductivity in amorphous heterogeneous nanocomposites}

\abbreviations{IR,NMR,UV}

\begin{document}

\begin{abstract}
  Pure amorphous solids are traditionally considered to set the lower bound of thermal conductivity due to their disordered atomic structure that impedes vibrational energy transport. However, the lower limits for thermal conductivity in heterogeneous amorphous solids and the physical mechanisms underlying these limits remain unclear. Here, we use equilibrium molecular dynamics to show that an amorphous SiGe nanocomposite can possess thermal conductivity substantially lower than those of the amorphous Si and Ge constituents. Normal mode analysis indicates that the presence of the Ge inclusion localizes vibrational modes with frequency above the Ge cutoff in the Si host, drastically reducing their ability to transport heat. This observation suggests a general route to achieve exceptionally low thermal conductivity in fully dense solids by restricting the vibrational density of states available for transport in heterogeneous amorphous nanocomposites. 
  
\end{abstract}

Low thermal conductivity materials are desired for a wide range of applications ranging from thermoelectric power generators\cite{poudel_high-thermoelectric_2008,biswas_high-performance_2012,zebarjadi_perspectives_2012,cahill_nanoscale_2014,zhao_panoscopic_2014,minnich_advances_2015} to thermopile detectors\cite{foote_high-performance_1998} .  Traditionally, amorphous materials are considered to set the lower limit of thermal conductivity due to the disordered atomic structure that impedes the formation of propatgating vibrations. \cite{cahill_thermal_1987,mcgaughey_phonon_2006,mizuno_beating_2015} While in crystals heat is carried by propagating lattice waves, or phonons, in amorphous solids the lack of a periodic atomic structure results in very different mechanisms for vibrational energy transport. 

Allen and Feldman introduced categories of vibrational modes in amorphous solids known as propagons, diffusons, and locons. \cite{allen_thermal_1989,allen_diffusons_1999} Propagons are propagating and delocalized phonon-like plane waves that typically possess long wavelengths compared to the interatomic spacing. Diffusons are modes that scatter over a distance less than their wavelength and thus transport heat as a random-walk. Locons are non-propagating and localized modes that are unable to transport heat in harmonic solids. \cite{larkin_thermal_2014,allen_diffusons_1999} 

This classification has been widely used to interpret experiments and calculations of transport in amorphous materials, particularly for pure a-Si. For instance, numerical works using equilibrium molecular dynamics (EMD) and lattice dynamics (LD) have attempted to determine the fraction of heat carried by each type of vibration. In their original work, Allen et al. reported that $\sim$20 \% of thermal conductivity of a-Si is from propagons ($\lesssim$ 3 THz) whereas the rest are from diffusons (3 - 17 THz) and none is from locons ($\gtrsim$ 17 THz). \cite{feldman_thermal_1993} He et al. reported that although only 3\% of the mode population is propagons, they transport up to 50\% of the heat due to their long propagation distances. \cite{he_heat_2011} Calculations by Larkin and McGaughey indicate that propagons have a lifetime scaling of $\omega ^{-2}$ which suggests that these modes are plane-wave-like and are propagating. \cite{larkin_thermal_2014} Wei and Henry have reported that frequency modes above $\sim$17 THz are highly localized and do not contribute to thermal conductivity using Green-Kubo modal analysis for a-Si.\cite{lv_direct_2016}.

Experimental works have qualitatively confirmed some of these predictions. \cite{cahill_thermal_1994,zink_thermal_2006,braun_size_2016} Sultan et al. reported that modification of the surface of an amorphous SiN membrane changes the thermal conductance of the membrane, indicating the importance of propagons for heat conduction. \cite{sultan_heat_2013} They estimated that propagons are responsible for $\sim$40-50 \% of thermal conductivity in amorphous SiN using kinetic theory. Braun et al. reported that diffusons are the dominant heat carriers for films of thickness less than 100 nm, while the propagon contribution is present in thicker films. \cite{braun_size_2016}.
 
Although pure amorphous solids are typically assumed to achieve the lower limit of thermal conductivity, some works have examined how this limit may be broken. In semi-crystalline solids, it is well known that thermal boundary resistance can result in exceptionally low thermal conductivity of composites \cite{minnich_modified_2007}. This effect has been exploited by Chiritescu et al. \cite{chiritescu_ultralow_2007} to achieve ultralow thermal conductivity in disordered WSe$_{2}$ nanolaminates below the minimum thermal conductivity predicted by the Cahill-Pohl model\cite{cahill_lower_1992}, although a recent theory work suggests that the experiments agree with this model if anisotropy is taken into account.\cite{chen_anisotropic_2015} Wingert et al. reported that crystalline silicon nanotubes with shell thicknesses as thin as 5 nm have a low thermal conductivity of 1.1 W/m-K, lower than that of the amorphous counterpart via a phonon softening effect. \cite{wingert_sub-amorphous_2015} Dechaumphai et al. experimentally observed an ultralow thermal conductivity of 0.33 $\pm$ 0.04 W/m-K at room temperature in amorphous multilayers made of Au and Si. \cite{dechaumphai_ultralow_2014} Computationally, Norouzzadeh et al. used MD to study the thermal conductivity of an a-SiGe alloy with different Ge content and observed thermal conductivity values below those of the constituent materials.\cite{norouzzadeh_thermal_2015}. Giri et al. used NEMD to examine the role of the interface of amorphous SiGe superlattices and amorphous Si/heavy-Si superlattices, concluding that increasing mass-mismatch in amorphous superlattices results in higher Kapitza resistances, which leads to lower thermal conductivity. \cite{giri_kapitza_2015} 

Although these works have suggested that thermal conductivities of heterogeneous amorphous solids below those of the pure constituents are achievable, key questions remain. Some of these works have interpreted their results with a phonon gas model, which is of questionable validity for diffusons and locons, and others have used the concept of thermal boundary resistance to explain their observations. In particular, the latter approach implicitly assumes that vibrational modes of the two solids composing the interface are well defined. However, if the inclusion in the nanocomposite is sufficiently small, the vibrational modes of the composite may not coincide with the vibrations of the pure materials. In this case, the nature of the vibrations in the composite solids and hence the lower limits of thermal conductivity in heterogeneous amorphous solids remain unclear. 

Here, we examine heat transport in amorphous SiGe nanocomposites consisting of a Ge inclusion in a Si host matrix. We find that these structures can possess thermal conductivities that are significantly smaller than those of the constituent materials, with the minimum thermal conductivity reaching as low as 32 \% of that of the amorphous Si host. Lattice dynamics analysis demonstrates that the presence of the Ge cluster drastically enhances localization of vibrational modes with frequency above the Ge cutoff in the Si host, leading to a remarkable decrease in thermal conductivity. These results demonstrate a mechanism for achieving remarkably low thermal conductivity in fully dense amorphous materials that may be useful for solid-state thermal insulation and highly sensitive thermopile detectors. 

\begin{figure}
\centering
\includegraphics[width=0.8\linewidth]{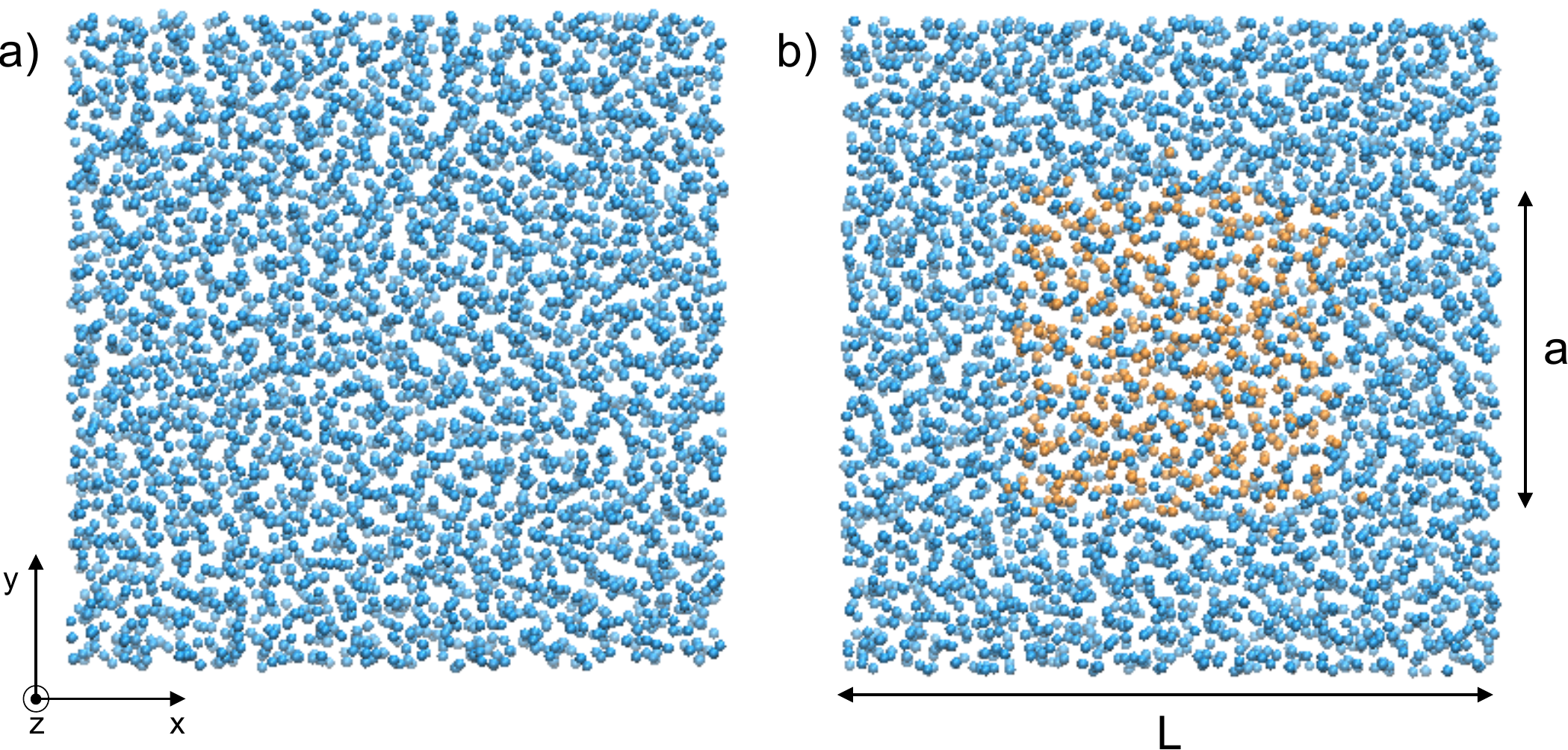}
\caption{4096-atom configurations of (a) amorphous silicon and (b) nanostructured amorphous silicon germanium. Blue atoms represent silicon and orange atoms represent germanium. The germanium cubic side length, $a$, varies from 10 \AA \space to the side length of the entire domain, $L$ = 46.4  \AA.}
\label{fig:domains}
\end{figure}

We calculated the thermal conductivity of amorphous Si and amorphous SiGe nanocomposites using equilibrium MD with the Stillinger-Weber (SW) interatomic potential.\cite{stillinger_computer_1985} The two types of structures studied are demonstrated in Figure \ref{fig:domains}. The atomic configuration consisting of 4096 atoms was provided by N. Mousseau and was generated from the modified Wooten-Winer-Weaire (WWW) algorithm.\cite{barkema_high-quality_2000}
For na-SiGe structures, a cubic domain in the middle of the structure with side length $a$ was replaced with heavier germanium atoms with appropriate coefficient changes in SW potential. SW potential coefficients for silicon and germanium interactions are described in Refs.\cite{stillinger_computer_1985,ding_molecular-dynamics_1986,laradji_structural_1995} The side length, $a$, was chosen to be 10, 20, 25, 30, 35, 40, and 46.4 \AA. These lengths represent 1, 10, 20, 35, 55, 82, and 100\% Ge fraction, respectively. Periodic boundary conditions were imposed for all the structures. The MD simulations were performed with Large-scale Atomic/Molecular Massively Parallel Simulator (LAMMPS)\cite{plimpton_fast_1995} with a time step of 0.5 fs. The simulation procedure began with an anneal at 1000 K for 20 ns using the NPT ensemble to reduce metastabilities.\cite{larkin_thermal_2014,he_heat_2011} We observed a decrease and plateau of the potential energy during the annealing process for each structure indicating a reduction of metastability.

Subsequently, the domain was quenched at a rate of 10 K/ps to 300 K and equilibrated in an NPT ensemble at 300 K for 20 ns to relax the structure to equilibrium pressure. Because volume and pressure fluctuate in MD simulations, we computed the average atom positions over the last 100 ps to ensure the domain was not under strain. The resulting mean pressure was on the order of 0.1 bar. This domain was then thermostatted in an NVT ensemble for 10 ns using a Nose-Hoover thermostat. After an additional NVE equilibration for 50 ps, the heat fluxes were computed for 1.6 ns in NVE ensemble.

We computed the thermal conductivity of the various structures using the Green-Kubo (GK) formalism, which relates the thermal conductivity to the heat current autocorrelation function by
\begin{equation}
k=\frac{V}{3k_BT^2}\int_{0}^{\infty}\langle\textbf{J}(t)\cdot\textbf{J}(0)\rangle dt\label{eqn:gk}
\end{equation}
where $k_B$ is the Boltzmann constant, $T$ is the temperature, $V$ is the system volume, $t$ is time, and $\textbf{J}$ is the heat flux. The angular brackets denote an ensemble average. The thermal conductivity calculations reported in this study are based on the average of the integrals of the heat current autocorrelation functions (HCACF) from 10 simulations. 

\begin{figure}
\centering
\includegraphics[width=1\linewidth]{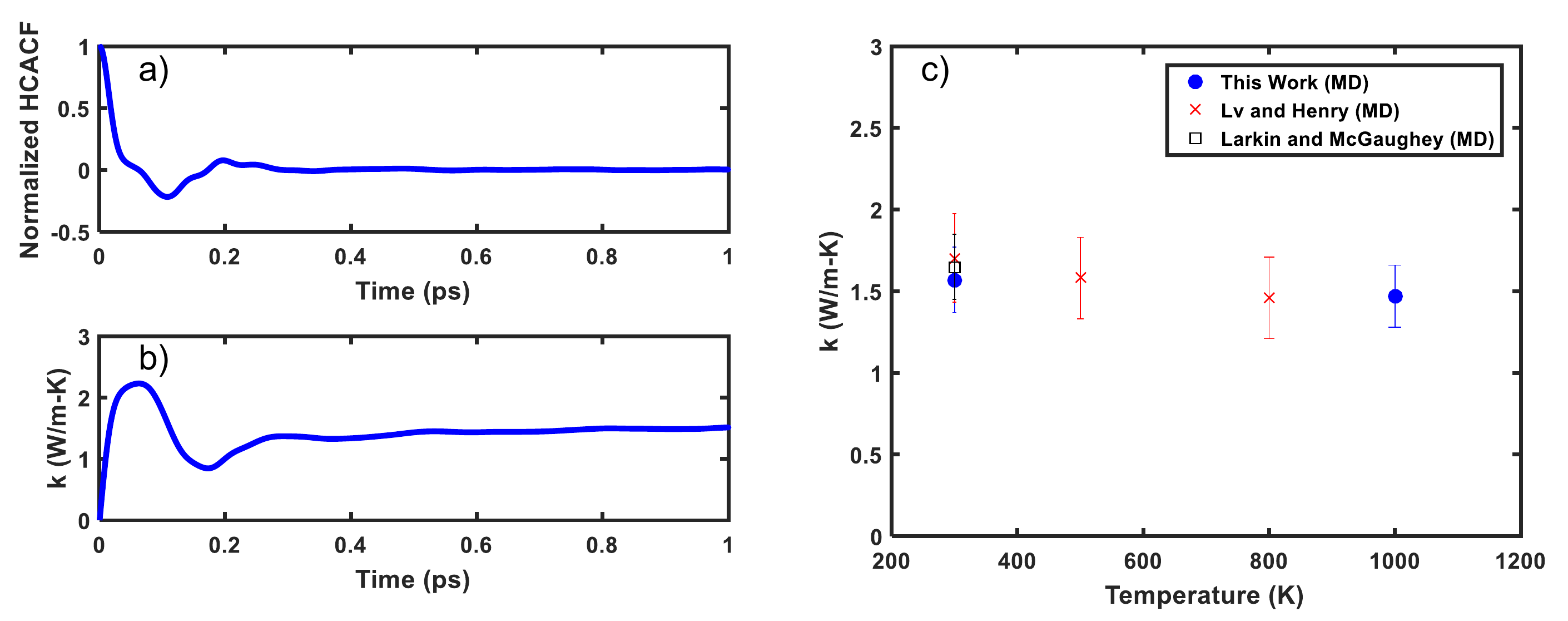}
\caption{(a) Normalized heat current autocorrelation function  (b) Thermal conductivity temporal profile calculated by Eq \ref{eqn:gk}. The thermal conductivity of a-Si is determined by taking the average between 5 and 20 ps. (c) Thermal conductivity versus temperature (blue circles) comparison with the works by  Larkin and McGaughey (black square) \cite{larkin_thermal_2014}, and Lv and Henry (red crosses) \cite{lv_direct_2016} utilizing 4096 atoms, SW potential, and GK formalism at temperatures from 300 K to 1000 K. No temperature dependence is observed. }
\label{fig:a-si}
\end{figure}

Figure \ref{fig:a-si}(a) shows the HCACF normalized by $\langle\textbf{J}(0)\cdot\textbf{J}(0)\rangle$ for a-Si. The autocorrelation function converges quickly to 0 in less than 0.5 ps. The resulting thermal conductivity obtained from the integral of the autocorrelation function versus integration time is depicted in Figure \ref{fig:a-si}(b). The thermal conductivity of a-Si is determined by taking the average between 5 and 20 ps. The thermal conductivity of a-Si with respect to temperature for 4096 atoms with SW potential is plotted in Figure \ref{fig:a-si}(c) and compared with works by Larkin and McGaughey \cite{larkin_thermal_2014} and Lv and Henry. \cite{lv_direct_2016}  For 300 K, thermal conductivity from this work is 1.55 $\pm$ 0.20 W/m-K which is in agreement with these works. Consistent with Ref\cite{lv_direct_2016}, weak temperature dependence of thermal conductivity is observed. Direct comparison to experimental results is difficult as thermal conductivity of a-Si varies significantly by the fabrication process, hydrogenation, heat treatment, and defects, but experimental thermal conductivity typically ranges from 1 to 6 W/m-K at room temperature.\cite{he_heat_2011,cahill_thermal_1994,zink_thermal_2006,liu_high_2009}

\begin{figure}
\centering
\includegraphics[width=0.5\linewidth]{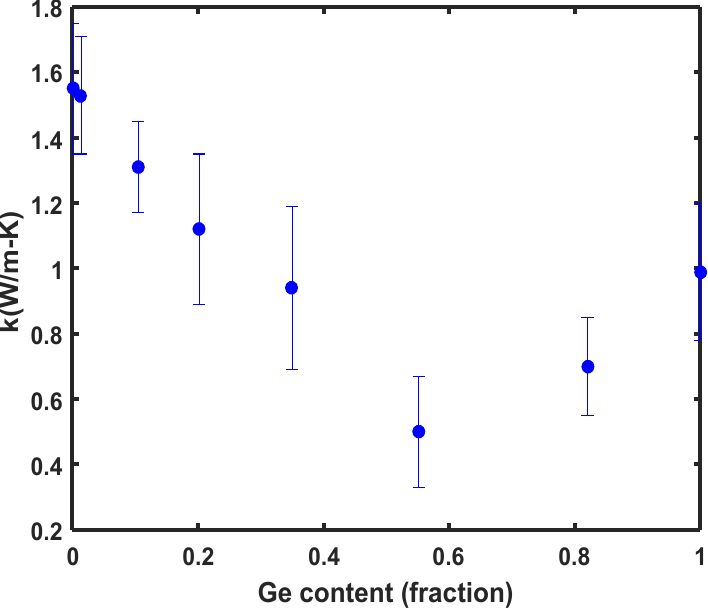}
\caption{Thermal conductivity of na-SiGe versus Ge content. The minimum thermal conductivity of 0.50 $\pm$ 0.17 W/m-K is observed with 55\% Ge content. }
\label{fig:gecontent}
\end{figure}

We now examine the thermal conductivity of na-SiGe versus Ge content, shown in Figure \ref{fig:gecontent}. Pure amorphous Si and Ge have thermal conductivities of 1.55 $\pm$ 0.20 W/m-K and 0.99 $\pm$ 0.21 W/m-K, respectively. Interestingly, we observe thermal conductivities substantially smaller than either of these values for na-SiGe composites with Ge content ranging from 35\% to 82\%, with the minimum thermal conductivity of 0.50 $\pm$ 0.17 W/m-K achieved with 55\% of Ge content. This value is less than a third of the original a-Si thermal conductivity and half that of a-Ge. The percentage decrease of thermal conductivity in na-SiGe is nearly twice that in a-Si/a-Ge superlattices by an NEMD study by Giri et al. utilizing SW potential despite similar geometry. \cite{giri_kapitza_2015}

\begin{figure}
	\centering
	\includegraphics[width=1\linewidth]{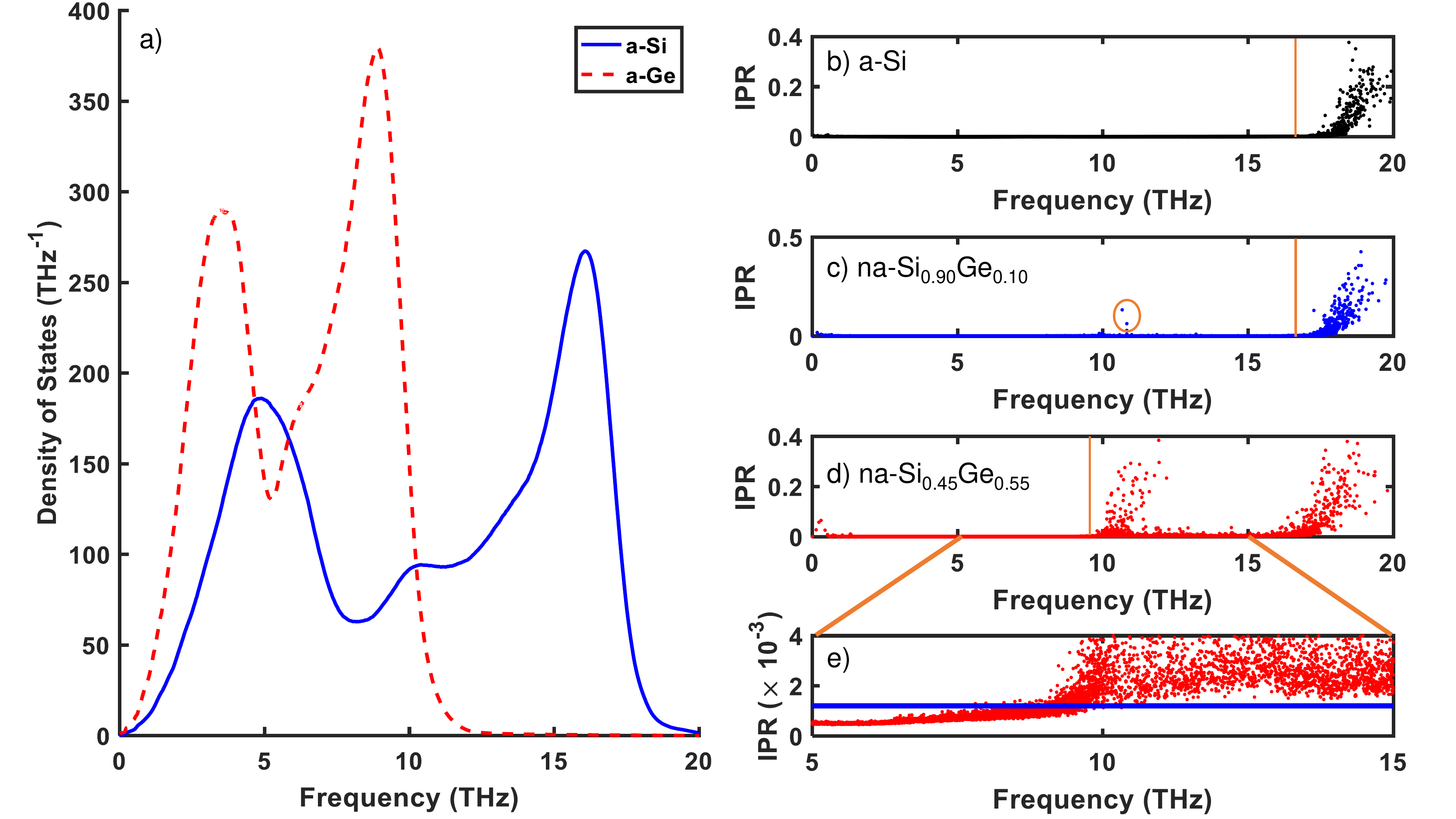}
	\caption{(a) The vibrational density of states of pure a-Si and a-Ge and inverse participation ratio (IPR) for (b) a-Si, (c) na-Si$_{0.90}$Ge$_{0.10}$, (d) na-Si$_{0.45}$Ge$_{0.55}$. (e) Zoomed-in view of na-Si$_{0.45}$Ge$_{0.55}$ for frequencies from 5 to 15 THz where above the bold line represents locons. Vibrational modes start to be localized at 9 THz and all become localized above 10 THz.}
	\label{fig:vdosiprinkscape}
\end{figure}

To understand the mechanism behind the reduction in thermal conductivity, we first examine the vibrational density of states (vDOS) of pure a-Si and a-Ge shown in Figure \ref{fig:vdosiprinkscape}(a). The vDOS is computed from
\begin{equation}
g(\omega) = \sum_{m=1}^{3N_{atom}}\delta(\omega-\omega_m) = \frac{1}{3k_BT} \int_{0}^{\infty} \sum_{n=1}^{N_{atom}}m_n \langle\textbf{v}(t)\cdot\textbf{v}(0)\rangle e^{i\omega t} dt\label{eq:vDOS}
\end{equation}
where $N_{atom}$ is the number of atoms, $T$ is the temperature, $m_n$ is the nth atom mass, $V_n(t)$ is the nth atom velocity at time t.\cite{dickey_computer_1969} The vDOS of a-Si and a-Ge is similar to that of c-Si and c-Ge with distinct peaks at certain frequencies.\cite{giri_kapitza_2015} Due to absence of strong anharmonicity, only weak vibrational interaction of Si and Ge atoms is expected for frequencies greater than the frequency cutoff of a-Ge of 10 THz. In other words, we expect the vibrational modes with frequencies exceeding 10 THz to be confined to a-Si.

 We confirm this hypothesis by first calculating the inverse participation ratio  (IPR), which is a measure of how many atoms participate in the motion of a particular eigenmode. The IPR is given by 
\begin{equation}
p_n^{-1}=\sum_{i}(\sum_{\alpha}e_{i \alpha, n}^{*}e_{i \alpha, n})^2\label{eq:IPR}
\end{equation}
where $e_{i \alpha, n}$ is the eigenvector component for atom $i$ in $\alpha$ direction for the mode $n$. \cite{bell_atomic_1970} The eigenvectors for each mode and atom are calculated by harmonic lattice dynamics in GULP \cite{gale_gulp:_1997} with relaxed structures from MD at 300 K. The IPR is defined so that it equals $1/N_{atom}$ if all atoms are participating, or 1 if the vibration is completely localized to one atom. Defining a specific IPR value that uniquely distinguishes locons is not possible, but vibrational modes with participation ratio less than 0.2 (corresponding to IPR greater than 0.0012 here) have been defined previously as localized modes.\cite{yang_extreme_2014,chen_remarkable_2010} We therefore define locons according to this convention.

Figures \ref{fig:vdosiprinkscape}(b)-(e) show the IPR for a-Si, na-Si$_{0.90}$Ge$_{0.10}$, na-Si$_{0.45}$Ge$_{0.55}$, and a zoomed-in view of the IPR of na-Si$_{0.45}$Ge$_{0.55}$ from 5 to 15 THz. The IPR for a-Si, Figure \ref{fig:vdosiprinkscape}(b), shows that locons are observed primarily over around 17 THz, consistent with prior works. \cite{allen_diffusons_1999,lv_direct_2016} As Ge atoms are introduced in the nanocomposite in na-Si$_{0.90}$Ge$_{0.10}$, we observe locons in the medium-frequency region around 10 THz. For na-Si$_{0.45}$Ge$_{0.55}$, all the vibrational modes above around 10 THz are localized. The corresponding locon mode fractions are 7\%, 9\%, and 31\% for a-Si, na-Si$_{0.90}$Ge$_{0.10}$, and na-Si$_{0.45}$Ge$_{0.55}$, respectively. In other words, na-Si$_{0.45}$Ge$_{0.55}$ has the lowest thermal conductivity and also more than 4 times the number of locons than a-Si, suggesting localized modes in Si are associated with the low thermal conductivity of the nanocomposite. We also note that vibrational modes with higher IPR than 0.0012 are present at low frequencies. We have verified that these modes are due to the finite size of the computational domain and disappear as the size of the system increases. 

We next confirm that these localized modes reside in silicon by calculating the local vibrational density of states, defined as \cite{feldman_vibrational_2004} \begin{equation}
D_i(\omega)=\sum_{n}\sum_{\alpha}e_{i \alpha, n}^{*}e_{i \alpha, n}\delta(\omega-\omega_n)\label{eq:LVDOS}
\end{equation}
where the sum is over Cartesian directions $\alpha$ and vibrational modes $n$ for atom $i$. Furthermore, the spatial distribution of energy can be described as\cite{chen_remarkable_2010} 
\begin{equation}E_i = \sum_{\omega}(n_{BE}+\frac{1}{2})\hbar\omega D_i(\omega)\label{eq:SDE}
\end{equation}
where $n_{BE}$ is the occupation number given by the Bose-Einstein distribution. We identify where the vibrational modes are localized by performing the sum only for vibrational modes that correspond to locons as identified by the IPR.   

\begin{figure}[h!]
\centering
\includegraphics[width=1\linewidth]{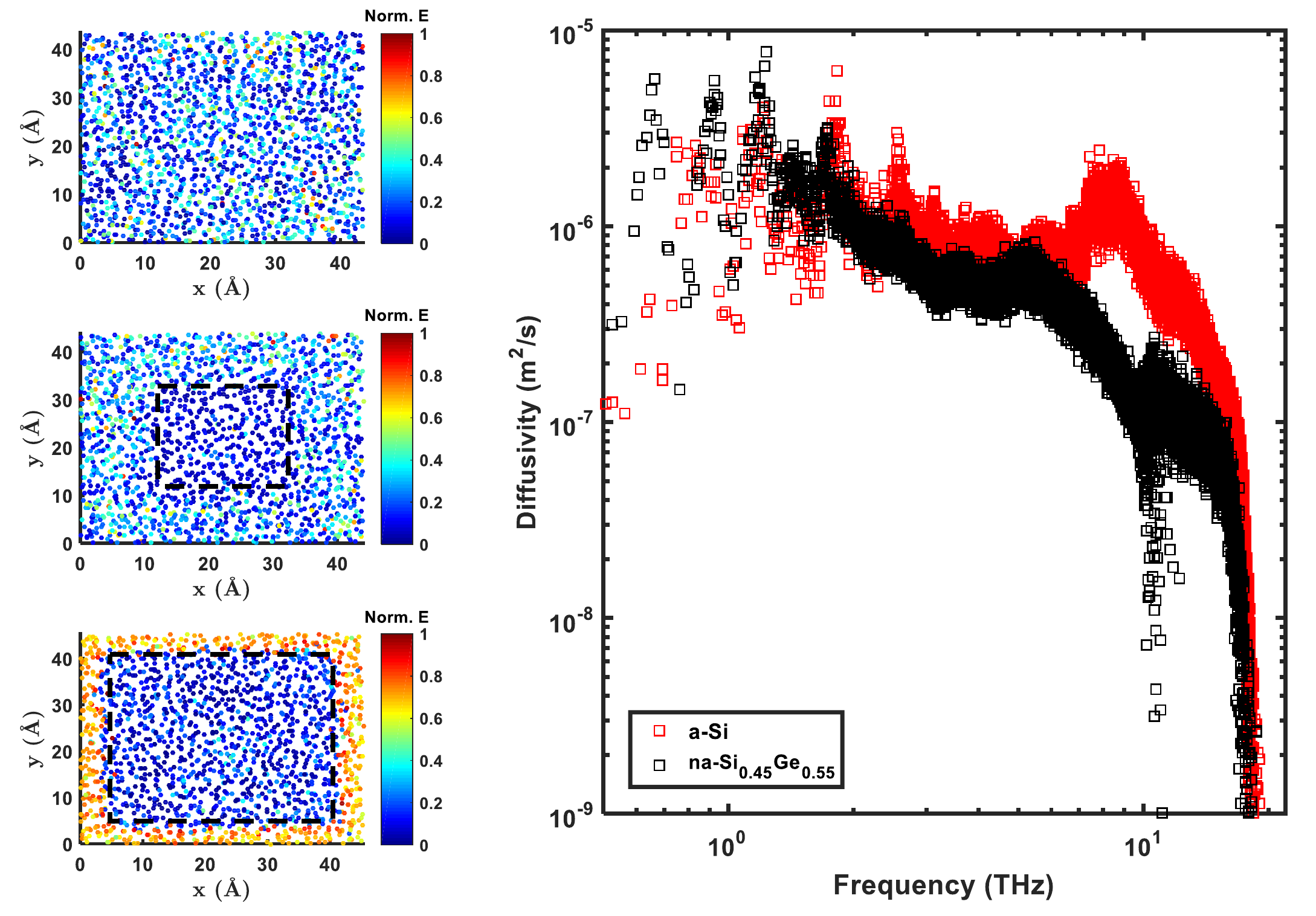}
\caption{Normalized spatial energy distribution of the cross section XY plane in the middle of z axis (a) a-Si, (b)na-Si$_{0.90}$Ge$_{0.10}$, and, (c) na-Si$_{0.45}$Ge$_{0.55}$. Individual circles in the figure represent atoms and dashed lines represent the boundaries between Si and Ge atoms. Red and blue atoms refer to localization and de-localization at these atoms, respectively. (d) Spectral thermal diffusivities of a-Si and na-Si$_{0.45}$Ge$_{0.55}$ versus mode frequency. Thermal diffusivities decrease significantly for vibrational modes with frequencies higher than 10 THz in na-Si$_{0.45}$Ge$_{0.55}$ compared to those in a-Si.}
\label{fig:lvdosdinkscape}
\end{figure}

The spatial energy distribution is shown in Figures \ref{fig:lvdosdinkscape}(a)-(c) for a-Si, na-Si$_{0.90}$Ge$_{0.10}$, and na-Si$_{0.45}$Ge$_{0.55}$, respectively. The distribution has been normalized by the maximum energy of an atom in the domain. We plot cross section x-y plane in the middle of z axis for clear visualization. It is apparent that for a-Si the spatial distribution of locons is randomly distributed. As Ge content is increased, however, we observe that locons are located in Si atoms. This result confirms that vibrational modes over around 10 THz are increasingly localized as Ge content grows and that these locons are indeed localized in a-Si atoms.

The drastic increase in locon population in na-SiGe suggests that the origin of the low thermal conductivity in na-SiGe is due to conversion of non-localized modes in a-Si to locons. To verify this hypothesis, we calculate the thermal diffusivities using the harmonic heat flux operator. The thermal conductivity of a solid is given by

\begin{equation}
k = \frac{1}{V}\sum_{i}C(\omega _i)D_{th}(\omega _i)\label{k}
\end{equation}
where V is the volume of structure, $C(\omega _i)$ is the specific heat, $D_{th}(\omega _i)$ is the thermal diffusivity of mode $\omega _i$, and the summation is over modes.
For diffusons under harmonic Allen-Feldman (AF) theory, the thermal diffusivity is calculated by 
\begin{equation}
D_{AF}(\omega _i)=\frac{\pi V^2}{\hbar ^2 \omega _i ^2}
\sum_{j \neq i} |S_{ij}|^2 \delta(\omega _i - \omega _j) 
\label{DAF}
 \end{equation}
 where $S_{ij}$ is the heat current operator in the harmonic approximation.\cite{allen_thermal_1993} Feldman et al. demonstrated that diffusivity calculations based on Peierls-Boltzmann theory (phonon gas model) for low frequency propagon modes coincide reasonably well with $D_{AF}$ in the low frequency range.\cite{feldman_vibrational_2004} Therefore, we calculate  $D_{AF}$ for all the vibrational modes for a-Si and na-Si$_{0.45}$Ge$_{0.55}$ as shown in Figure \ref{fig:lvdosdinkscape}(d). We observe that, for vibrational modes defined as locons by IPR ($\gtrsim$ 17 THz for a-Si and $\gtrsim$ 10 THz for na-Si$_{0.45}$Ge$_{0.55}$), the thermal diffusivities decrease significantly. For vibrational modes with frequencies between 10 THz to 17 THz, we observe an order of magnitude decrease in diffusivity from a-Si to na-Si$_{0.45}$Ge$_{0.55}$, contributing to the decrease in thermal conductivity. For low frequency propagating modes $\lesssim$ 2 THz,  no apparent changes in diffusivity occur among different structures, and we are unable to conclude how propagons with frequencies less than 1 THz are affected in the nanocomposite due to limitations in the size of the domain.
 
 The results suggest a simple explanation for the low thermal conductivity of the nanocomposite. In a-Si, nearly the full vibrational spectrum contributes to heat conduction as indicated by the calculated thermal diffusivities and associated small locon population. In the nanocomposite, diffusons with frequencies above the soft Ge cutoff frequency become localized, impeding their ability to transport heat. In effect, the soft inclusion restricts the vibrational spectrum available to conduct heat because many Si vibrational modes are not supported in the inclusion. 
 
 Another interesting consideration is why the thermal conductivity of the composite is less than the intrinsic thermal conductivity of the amorphous Ge. The explanation can again be identified from the density of vibrational states. Although the a-Ge has a lower cutoff frequency than a-Si, its density of states is the same as that as a-Si because the atomic number densities are identical. However, in the nanocomposite, only a fraction of the modes in Si are able to conduct heat; therefore, the density of states with non-negligible thermal diffusivities in the nanocomposite are less than in a-Ge. As a result, the thermal conductivity of the composite may be lower than those of both the stiff host and softer inclusion.
 
Many prior works have interpreted thermal conductivity reductions in amorphous or disordered heterogeneous solids using the concept of thermal boundary resistance between the adjacent layers \cite{cahill_nanoscale_2014,dechaumphai_ultralow_2014,giri_kapitza_2015}.  However, this interpretation relies on the vibrational mode properties of individual constituents separately. Our analysis shows that the vibration mode characters change drastically from a-Si to na-SiGe, suggesting that the thermal boundary resistance is not a well-defined concept in the amorphous nanocomposites studied here as the vibrational modes of the constituent materials cannot be separated. Instead, it is the change in character of the overall vibrational modes of the composite that leads to the low thermal conductivity.

In summary, we have studied thermal transport in amorphous heterogeneous nanocomposites using molecular dynamics and lattice dynamics. We find that the thermal conductivity of na-Si$_{0.45}$Ge$_{0.55}$ is substantially lower than that of both constituent materials due to the localization of vibrational modes in the stiff host a-Si with frequencies exceeding the cutoff of the soft inclusion. This observation suggests a general route to achieve exceptionally low thermal conductivity in fully dense amorphous solids by restricting the vibrational density of states for transport in heterogeneous nanocomposites.

\begin{acknowledgement}

This work was supported by the Samsung Scholarship, NSF CAREER Award CBET 1254213, and Boeing under Boeing-Caltech Strategic Research and Development Relationship Agreement and used the Extreme Science and Engineering Discovery Environment (XSEDE), which is supported by National Science Foundation grant number ACI-1053575. The authors thank Normand Mousseau for providing the atomic positions from the WWW algorithm, and Andrew Robbins, Benoit Latour, Wei Lv, and Asegun Henry for useful discussions. 

\end{acknowledgement}


\providecommand{\latin}[1]{#1}
\makeatletter
\providecommand{\doi}
{\begingroup\let\do\@makeother\dospecials
	\catcode`\{=1 \catcode`\}=2\doi@aux}
\providecommand{\doi@aux}[1]{\endgroup\texttt{#1}}
\makeatother
\providecommand*\mcitethebibliography{\thebibliography}
\csname @ifundefined\endcsname{endmcitethebibliography}
{\let\endmcitethebibliography\endthebibliography}{}

\end{document}